\documentclass[seceq]{ptptex}
\usepackage{graphicx}
\begin{document}

\markboth{A. Gendiar, R. Krcmar, and T. Nishino}{
Spherical Deformations for 1D Quantum Systems}

\title{Spherical Deformation for one-dimensional Quantum Systems}

\author{Andrej \textsc{Gendiar}$^{1,2}$, Roman \textsc{Krcmar}$^1$, and 
Tomotoshi \textsc{Nishino}$^{2,3}$}

\inst{$^1$ Institute of Electrical Engineering,
Slovak Academy of Sciences, D\'{u}bravsk\'{a} cesta 9, SK-841~04, Bratislava, Slovakia\\
$^2$ Institute for Theoretical Physics C, RWTH University Aachen, D-52056 Aachen, Germany\\
$^3$ Department of Physics, Graduate School of Science, Kobe University,
Kobe 657-8501, Japan}


\abst{
System-size dependence of the ground-state energy $E^N_{~}$ is investigated
for $N$-site one-dimensional (1D) quantum systems with open boundary 
condition, where the interaction strength decreases towards the 
both ends of the system. For the spinless Fermions on the 1D lattice we have
considered, it is shown that the finite-size correction to
the energy per site, which is defined as $E^N_{~} \! / N - 
\lim_{N \rightarrow \infty}^{~} E^N_{~} \! / N$, 
is of the order of $1 / N^2_{~}$ when the reduction factor of the
interaction is expressed by a sinusoidal function. We discuss the origin of this
fast convergence from the view point of the spherical geometry.
}



\maketitle

\section{Introduction}

A purpose of numerical studies in condensed matter physics is to
obtain bulk properties of systems in the thermodynamic limit. In
principle numerical methods are applicable to systems with {\it finite}
degrees of freedom, and therefore occasionally it is impossible to treat {\it infinite}
system directly. A way of estimating the thermodynamic limit is 
to study finite-size systems, and subtract the finite-size corrections
by means of extrapolation with respect to the system size.~\cite{FSS,FSS2}.

As an example of extensive functions, which is essential for bulk
properties, we consider the ground state energy $E^N_{~}$ of $N$-site 
one-dimensional (1D) quantum systems. In this article we focus on the convergence
of energy per site $E^N_{~} / N$ with respect to the system size $N$.
In order to clarify the discussion, we specify the form of lattice 
Hamiltonian 
\begin{equation}
\hat H = 
\sum_\ell^{~} \, \hat h_{\ell, \ell+1}^{~} + 
\sum_\ell^{~} \, \hat g_\ell^{~} \, ,
\end{equation}
which contains on-site terms $\hat g_\ell^{~}$ and nearest neighbor 
interactions $\hat h_{\ell, \ell+1}^{~}$. We assume that the operator 
form of $\hat h_{\ell, \ell+1}^{~}$ and $\hat g_\ell^{~}$ are independent 
of the site index $\ell$, which means that $\hat H$ is translationally
invariant in the infinite $N$ limit. It is possible to include 
$\hat g_\ell^{~}$ into $\hat h_{\ell, \ell+1}^{~}$ by the redefinition
\begin{equation}
\hat h_{\ell, \ell+1}^{~} + 
\frac{ \hat g_\ell^{~} + \hat g_{\ell+1}^{~} }{2} 
\,\,\,\, \rightarrow \,\,\,\,
\hat h_{\ell, \ell+1}^{~} \, ,
\end{equation}
and therefore we group $\hat g_\ell^{~}$ with $\hat h_{\ell, \ell+1}^{~}$ 
as shown in Eq.~(1$\cdot$2) if it is convenient.
A typical example of such $\hat H$ is the spin Hamiltonian of the Heisenberg chain 
\begin{equation}
\hat H = 
J \, \sum_\ell^{~} \, \hat {\bf S}_\ell^{~} \cdot \hat {\bf S}_{\ell+1}^{~} - 
B \, \sum_\ell^{~} \, \hat S_\ell^{Z} \, ,
\end{equation}
where $\hat {\bf S}_\ell^{~}$ represents the spin operator at $\ell$-th site,
and $\hat S_\ell^{Z}$ its $Z$-component. 
The parameters $J$ and $B$ are, respectively, the neighboring interaction 
strength and the external magnetic field. 
In this case $\hat h_{\ell, \ell+1}^{~}$ and $\hat g_\ell^{~}$ are, respectively,
$J \, \hat {\bf S}_\ell^{~} \cdot \hat {\bf S}_{\ell+1}^{~}$ and 
$- B \hat S_\ell^{Z}$. 
If the chain is infinitely long, $\hat H$ in Eq.~(1$\cdot$3) 
is translational invariant, and the ground state $| \Psi_0^{~}
\rangle$ is uniform when there is no symmetry breaking. 
For example, the bond-energy
$J \, \langle  \, \hat {\bf S}_\ell^{~} \cdot \hat {\bf S}_{\ell+1}^{~} \rangle 
= J \, \langle \Psi_0^{~} | \,  \hat {\bf S}_\ell^{~} \cdot \hat {\bf S}_{\ell+1}^{~} 
\, | \Psi_0^{~} \rangle$ of the integer-spin Heisenberg chain is
independent on $\ell$.

This homogeneous property of the system is violated if 
only a part of the interactions $\hat h_{1,2}^{~}$, $\hat h_{2,3}^{~}$,
$\ldots$, and $\hat h_{N-1, N}^{~}$ is present, and the rest does not exist. 
In other words, if we consider an $N$-site open boundary system defined by
the Hamiltonian
\begin{equation}
\hat H_{\rm Open}^{~} = 
\sum_{\ell=1}^{N-1} \, \hat h_{\ell, \ell+1}^{~} + 
\sum_{\ell=1}^N \, \hat g_\ell^{~} \, ,
\end{equation}
the ground state $| \Psi_0^{~} \rangle$ is normally non-uniform.
As a result the expectation values $\langle \hat h_{\ell, \ell+1}^{~}
\rangle$ and $\langle \hat g_\ell^{~} \rangle$ are  position 
dependent, especially near the boundary of the system.
The ground state energy $E_{~}^{N}$ of this $N$-site 
system is normally not proportional to the system size $N$, which
shows the presence of boundary energy correction. 
Such a finite-size effect is non-trivial when the system is gapless, 
as observed in the $S = 1/2$ Heisenberg spin chain~\cite{White2}. 

In case that we are interested in the bulk property of the system,
it is better to reduce the boundary effect as rapidly as possible. 
For this purpose Veki\'c and White introduced a sort of 
smoothing factor $A_\ell^{~}$ to the Hamiltonian
\begin{equation}
\hat H_{\rm Smooth}^{~} = 
\sum_{\ell=1}^{N-1} A_\ell^{~} \, 
\left( \hat h_{\ell, \ell+1}^{~} + \frac{ \hat g_\ell^{~} + \hat g_{\ell+1}^{~}}{2}
\right) \, ,
\end{equation}
where $A_\ell^{~}$ is almost unity deep inside the system and
decays to zero near the both boundaries of the
system~\cite{Vekic}. The factor $A_\ell^{~}$ is adjusted so that
the boundary effect disappears rapidly with respect to the
distance from the boundary. A simplest parametrization is to reduce only 
$A_1^{~}$ and $A_{N-1}^{~}$ from unity, leaving other factors equal to
unity. This simple choice of $A_\ell^{~}$ is often used for calculations of
the Haldane gap~\cite{Huse}.

As an alternative approach, Ueda and Nishino recently introduced
the {\it hyperbolic deformation}, which is characterized by the
non-uniform Hamiltonian
\begin{equation}
\hat H_{\rm Hyp.}^{~} = 
\sum_{\ell = 1}^{N-1} 
\cosh \left( \lambda \, \frac{2 \ell - N - 1}{2} \right) \,\, 
\left( \hat h_{\ell, \ell+1}^{~} + \frac{ \hat g_\ell^{~} + \hat g_{\ell+1}^{~}}{2} \right)
\, ,
\end{equation}
where $\lambda$ is a small positive
constant of the order of $0.01 \sim 0.1$~\cite{hyp,hyp2}.
As long as the form of the Hamiltonian is concerned, $\hat H_{\rm Hyp.}^{~}$ 
can be regarded as a special case of $\hat H_{\rm Smooth}^{~}$ in 
Eq.~(1$\cdot$5) with $A_\ell^{~} = 
\cosh \left( \lambda \, {\textstyle \frac{2 \ell - N - 1}{2}} \right)$. 
But in the scheme of hyperbolic deformation, the factor 
$\cosh \left( \lambda \, {\textstyle \frac{2 \ell - N - 1}{2}} \right)$ is
an increasing function of $| ( 2 \ell - N - 1 ) / 2  |$, 
and therefore the boundary effect is in principle enhanced. This enhancement works
uniformly for most of the lattice sites, and the expectation value
$
\langle h_{\ell, \ell+1}^{~} \rangle =
\langle \Psi_0^{~}  | \, h_{\ell, \ell+1}^{~} \, | \Psi_0^{~}  \rangle
$
for the ground state $| \Psi_0^{~} \rangle$ becomes nearly
independent on $\ell$ for most of the bonds. After obtaining the
expectation value $\langle h_{\ell, \ell+1}^{~} \rangle$ at the center of
the system for several values of the deformation parameter $\lambda$, 
one can perform an extrapolation
towards $\lambda = 0$ to get the energy per site of the undeformed
system. Such an extrapolation is possible since the hyperbolic deformation
has an effect of decreasing the correlation length of the system.

The hyperbolically deformed system is closely related to classical
lattice models on the hyperbolic plane with a constant and negative
curvature.~\cite{penta1,penta2,penta3,Baek,Sausset,Angles,Madras,
Chris,Shima,Hasegawa,Rietman,Doyon} In this article we imagine the 
case of a positive constant curvature, where the classical lattice 
models are on a sphere. The corresponding quantum Hamiltonian 
can be written as
\begin{equation}
\hat H_{\rm Sph.}^{~} = \sum_{\ell = 1}^{N-1} 
\sin \frac{\ell \pi}{N} \, 
\left( \hat h_{\ell, \ell+1}^{~} + \frac{ \hat g_\ell^{~} + \hat g_{\ell+1}^{~}}{2} \right)
\, ,
\end{equation}
where $A_\ell^{~} = \sin( \ell \pi / N )$ 
decreases to zero toward the system boundary. 
We call such a modification of the bond strength
as the {\it spherical deformation}, and consider $N$
as the system size. We analyze the ground state 
$| \Psi_0^{~} \rangle$ and the ground-state energy $E^N_{~}$ 
of this deformed Hamiltonian for the case of spinless 
free Fermions on the lattice. We  find that the difference
\begin{equation}
\frac{E^N_{~}}{N} - \lim_{N \rightarrow \infty}^{~} \frac{E^N_{~}}{N} \, ,
\end{equation}
which is the finite-size correction included in the energy per site $E^N_{~} / N$,
is of the order of $1 / N^2_{~}$. Note that this $1/N^2_{~}$ dependence is
the same as observed for the system with periodic boundary
conditions, described by the Hamiltonian 
\begin{equation}
\hat H_{\rm Periodic}^{~} = \sum_{\ell = 1}^{N-1} 
\left( \hat h_{\ell, \ell+1}^{~} + \frac{ \hat g_\ell^{~} + \hat g_{\ell+1}^{~}}{2} \right) + 
\left( \hat h_{N,1}^{~} + \frac{ \hat g_N^{~} + \hat g_{1}^{~}}{2} \right)
\, .
\end{equation}
In a certain sense, the spherically deformed system does not contain system boundary.

Structure of this article is as follows. In the next section we
introduce a spinless free Fermion model on 1D lattice. For tutorial purpose, the
finite-size effect is reviewed for systems with open and periodic
boundary conditions. In Sec.~3 we show our numerical results
obtained from the diagonalization of the spherically deformed
Hamiltonian $\hat H_{\rm Sph.}^{~}$ in Eq.~(1$\cdot$7). In Sec.~4 we consider
geometrical meaning of the spherical deformation by way of the Trotter
decomposition applied to the deformed Hamiltonian. We also consider a 
continuous limit, where the lattice spacing becomes zero. 
We summarize the obtained results in the last section.

\section{Energy corrections in the free fermion system}

As an example of 1D quantum systems, we consider the spinless free 
Fermions on the 1D lattice. The Hamiltonian is defined as
\begin{equation}
\hat H = -t \, \sum_\ell^{~} \left(
\hat c_{\ell}^{\dagger} \hat c_{\ell+1}^{~} + 
\hat c_{\ell+1}^{\dagger} \hat c_{\ell}^{~} \right) -
\mu \, \sum_\ell^{~} 
\hat c_{\ell}^{\dagger} \hat c_{\ell}^{~} \, ,
\end{equation}
where $t$ and $\mu$ are, respectively, the hopping parameter and 
the chemical potential. For simplicity we set $\mu = 0$ and treat the 
half-filled state in this Section when $\mu$ is not explicitly shown.
As a preparation for the spherical deformation, let us observe the
ground state properties of the above Hamiltonian, when open or
periodic boundary conditions are imposed for finite-size systems at 
half filling.

First we consider the $N$-site system with open boundary 
conditions, where the Hamiltonian is written as
\begin{equation}
\hat H^{~}_{\rm O}= -t \, \sum_{\ell = 1}^{N-1} \left(
\hat c_{\ell}^{\dagger} \hat c_{\ell+1}^{~} + 
\hat c_{\ell+1}^{\dagger} \hat c_{\ell}^{~} \right) \, .
\end{equation}
Since there is no interaction, 
the one-particle eigenstate $| \psi_m^{~} \rangle$ represented by the wave function
\begin{equation}
\langle 0 | \hat c_\ell^{~} | \psi_m^{~} \rangle = 
\psi^{~}_{m}( \ell ) = \sqrt{\frac{2}{N+1}} \,
\sin \frac{m \pi  \ell}{N+1} 
\end{equation}
is essential for the ground-state analysis,
where $m$ is the integer within the range $1 \le m \le N$.
The corresponding one particle energy is
\begin{equation}
\varepsilon_{m}^{~} = - 2 t \, \cos \frac{m \pi }{N+1}  \, ,
\end{equation}
and the ground-state energy at half filling is 
obtained by summing up all the negative eigenvalues.
Assuming that $N$ is even, the ground-state energy is 
obtained as
\begin{equation}
E_{\rm O}^{N}  =
\sum_{m=1}^{N/2} \varepsilon_{m}^{~} =
t \left[ 1 - \left( \sin \frac{\pi / 2}{N+1}  \right)^{-1}_{~} \right] \, 
\end{equation}
after a short calculation. Expanding the r.h.s. with respect to $N$, 
one finds the asymptotic form
\begin{equation}
\frac{E_{\rm O}^{N}}{N}  \sim - \frac{2}{\pi} t +
\frac{t}{N} \left( \frac{2}{\pi} - 1 \right) \, .
\end{equation}
Compared with the energy per site in the thermodynamic limit
\begin{equation}
\lim_{N \rightarrow \infty}^{~} \frac{E_{\rm O}^{N}}{N} = 
\frac{1}{\pi} \int_0^{\pi/2} - 2 t \, \cos k \, d k = -
\frac{2}{\pi} t \, ,
\end{equation}
it is shown that
the finite-size correction to the energy per site (or even to the
energy per bond) is of the order of $1/N$.

The $N$-dependence of the energy correction changes if we impose 
the periodic boundary conditions, where the Hamiltonian is given by
\begin{equation}
\hat H^{~}_{\rm P}= -t \, \sum_{\ell = 1}^{N-1} \left(
\hat c_{\ell}^{\dagger} \hat c_{\ell+1}^{~} + 
\hat c_{\ell+1}^{\dagger} \hat c_{\ell}^{~} \right)
- t \left(
\hat c_{N}^{\dagger} \hat c_{1}^{~} + 
\hat c_{1}^{\dagger} \hat c_{N}^{~} \right) \, .
\end{equation}
In this case, the one-particle wave function is the plane wave
\begin{equation}
\psi^{~}_{m}( \ell ) = \sqrt{\frac{1}{N}} \, \exp\left[
 i \frac{2 m \pi  (\ell - 1)}{N} \right]
\, ,
\end{equation}
where $m$ is an integer that satisfies $- N/2 + 1 < m \le N/2$.
The corresponding one-particle energy is
\begin{equation}
\varepsilon_{m}^{~} = - 2 t \, \cos \frac{2 m \pi }{N} \, .
\end{equation}
If $N$ is a multiple of four, the ground state energy at half
filling is calculated as
\begin{equation}
E_{\rm P}^{N}  =
\sum_{m=-N/4+1}^{N/4} \varepsilon_{m}^{~} =
- 2 t \, \cot \frac{\pi}{N}  \, .
\end{equation}
Thus, the finite-size correction to the energy per site
\begin{equation}
\frac{E_{\rm P}^{N} }{N} - \left( - \frac{2}{\pi} t \right) 
= - \frac{2 t}{N} \, \cot \frac{\pi}{N} + \frac{2 t}{\pi}
\,\, \sim \,\, \frac{2 \pi t}{3N^2_{~}}
\end{equation}
is of the order of $1/{N^2_{~}}$.

As verified in the above calculations, the finite-size correction
to the energy per site $E^N_{~} / N$ decreases faster for the system with
the periodic boundary conditions than with the open boundary conditions.
Regardless of this fact, the open boundary systems are often chosen in
numerical studies by the density matrix renormalization group
(DMRG) method~\cite{White2,White1,Pe,Scholl} because of the
simplicity in numerical calculation. It should be noted that for those systems 
that exhibits incommensurate modulation, the open boundary condition is
more appropriate than the periodic boundary condition.
Thus, it will be convenient if there is a way of decreasing the finite-size 
correction to  $E^N_{~} / N$  as fast as $1 / N^2_{~}$ also for the open 
boundary systems.

\section{Spherical deformation}

\begin{figure}
\centerline{\includegraphics[width=7.2cm,clip]{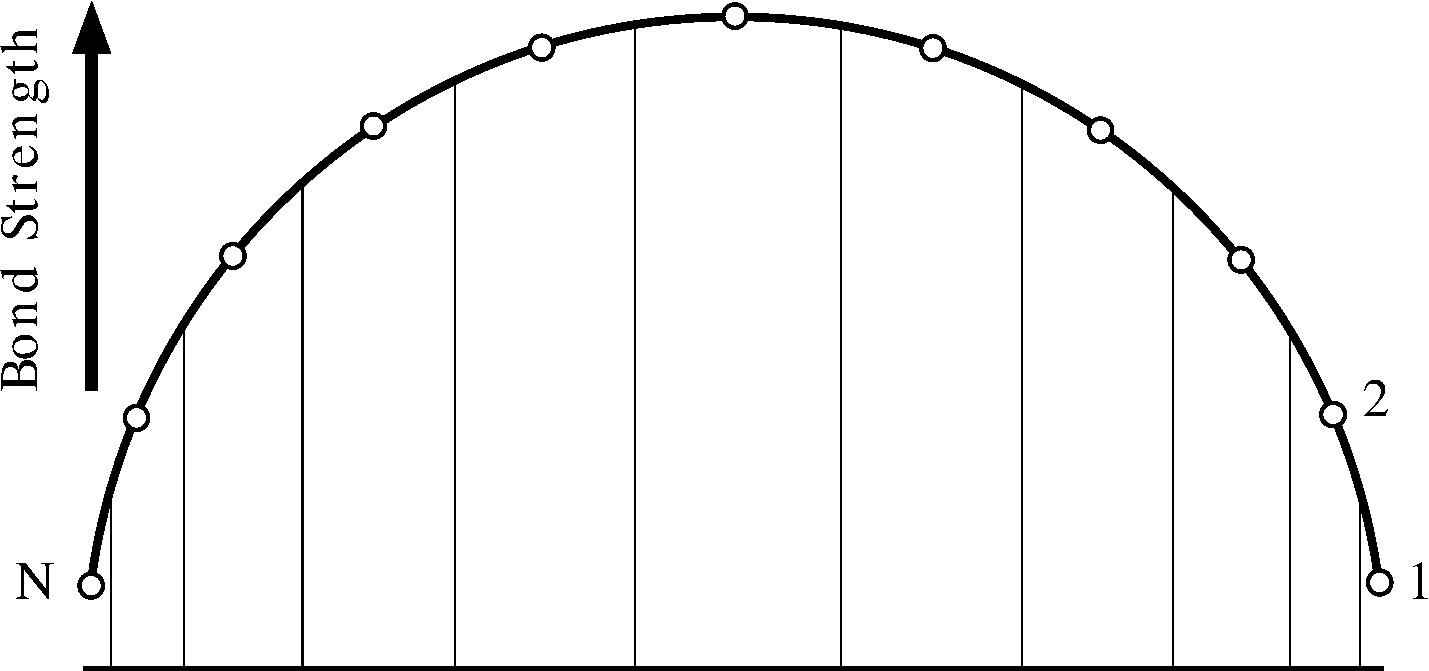}}
\caption{ 
A spherically deformed lattice, which contains $(N \! = \! 11)$-sites, drawn on the upper half 
of the circumference. Open circles denote lattice sites, where the angle of the
$\ell$-th site is $\theta_\ell^{~} = ( \ell - \frac{1}{2} ) \pi / N$ for 
$\ell = 1, \, 2, \, \ldots, \, N$. The length of the vertical 
line shows the
relative strength $\sin( \ell \pi / N )$ of the bond drawn by the thick arc between 
$\ell$-th and $(\ell+1)$-th sites.}
\label{f1}
\end{figure}

We first consider the $N$-site open boundary system described by the
Hamiltonian
\begin{equation}
\hat H^{~}_{\rm S} = -t \, \sum_{\ell = 1}^{N-1}
\sin  \frac{ \ell \pi   }{N} \, 
\left( 
\hat c_{\ell}^{\dagger} \hat c_{\ell+1}^{~} + 
\hat c_{\ell+1}^{\dagger} \hat c_{\ell}^{~} \right) \, .
\end{equation}
Compared with the undeformed Hamiltonian $\hat H^{~}_{\rm O}$ in
Eq.~(2$\cdot$2), the strength of the hopping term is scaled by the factor
$A_\ell^{~} = \sin( \ell \pi / N )$, which
decreases towards the system boundary as shown in Fig.~1.
For a geometrical reason which we discuss in the next
section, we call the modification from $\hat H^{~}_{\rm O}$ to $\hat H^{~}_{\rm
S}$ as the {\it spherical deformation}. We regard $N$, the number of sites on the upper
half of the circumference shown in Fig.~1, as the system size. 

Let us observe the $N$ dependence 
of the ground-state energy at half filling, where 
$n_\ell^{~} = \langle {\hat c}_\ell^\dagger {\hat c}_\ell^{~} \rangle 
= 1/2$ is satisfied by the particle-hole symmetry.
So far we have not obtained the analytic form of the one-particle wave function
$\psi^{~}_{m}$, except for the zero-energy state,
and the corresponding one-particle eigenvalue $\varepsilon^{~}_{m}$
for the deformed Hamiltonian $\hat H_{\rm S}^{~}$.
We therefore calculate them numerically by diagonalizing $\hat H^{~}_{\rm S}$ in 
the one-particle subspace. We then obtain the expectation value 
$\langle \hat c_\ell^{\dagger} \hat c_{\ell+1}^{~} +
\hat c_{\ell+1}^{\dagger} \hat c_{\ell}^{~} \rangle$ and
the ground state energy $E_{\rm S}^N$ at half filling.
In the following numerical calculations, we set $t$ as the unit of the energy.

\begin{figure}
\centerline{\includegraphics[width=8.2cm,clip]{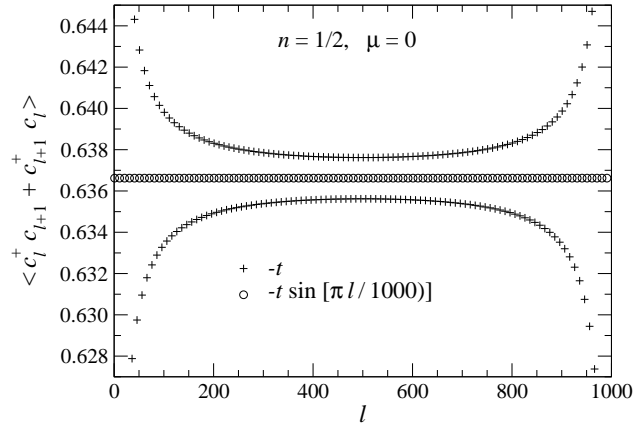}}
\caption{The circles shows the
expectation value $\langle \hat c_\ell^{\dagger}
\hat c_{\ell+1}^{~} + \hat c_{\ell+1}^{\dagger} \hat c_{\ell}^{~} \rangle$ of the
spherically deformed lattice Fermion model defined by ${\hat H}_{\rm S}^{~}$ 
when $N = 1000$. For comparison, we also plot the same expectation value for
 the undeformed case defined by ${\hat H}_{\rm O}^{~}$ by the cross marks.}
\label{f2}
\end{figure}

Figure~2 shows $\langle
\hat c_\ell^{\dagger} \hat c_{\ell+1}^{~} + \hat c_{\ell+1}^{\dagger}
\hat c_{\ell}^{~} \rangle$ of the ground state 
when $N = 1000$. For comparison, we also show the same
quantity obtained by the undeformed Hamiltonian $\hat H^{~}_{\rm
O}$ of the same system size. As it is observed, the spherical 
deformation suppresses the
position dependence in $\langle \hat c_\ell^{\dagger} \hat c_{\ell+1}^{~} +
\hat c_{\ell+1}^{\dagger} \hat c_{\ell}^{~} \rangle$. In this sense 
we can say that the ground state of
$\hat H^{~}_{\rm S}$ is more uniform than that of $\hat H^{~}_{\rm O}$. 

One expects that the ground state energy $E^N_{\rm S}$, which is
the sum of negative one-particle eigenvalues
\begin{equation}
E^N_{\rm S} = \sum_{m = 1}^{N/2} \varepsilon^{~}_{m} = 
- t \, \sum_{\ell = 1}^{N-1} 
\sin \frac{ \ell \pi }{N} \, 
\langle \hat c_\ell^{\dagger} \hat c_{\ell+1}^{~} +
\hat c_{\ell+1}^{\dagger} \hat c_{\ell}^{~} \rangle \, ,
\end{equation}
is nearly proportional to the sum of the bond strength
\begin{equation}
B^N_{~} = \sum_{\ell = 1}^{N-1}
\sin \frac{ \ell \pi }{N} 
= \cot \frac{\pi}{2N}  \, .
\end{equation}
It is also expected that the ratio $E^N_{\rm S} / B^N_{~}$ rapidly converges to
$- 2 t / \pi$, which is the expectation value 
$\langle \hat c_\ell^{\dagger} \hat c_{\ell+1}^{~} +
\hat c_{\ell+1}^{\dagger} \hat c_{\ell}^{~} \rangle$ in the thermodynamic limit.
Figure~3 shows $E^N_{\rm S} /
B^N_{~}$ and $E^N_{\rm P} / N$ with respect to $1 / N^2_{~}$.
Obviously, the finite-size corrections $E^N_{\rm P} / N + 2 t / \pi$ 
and $E^N_{\rm S} / B^N_{~} + 2 t / \pi$ 
are nearly proportional to $1 / N^2_{~}$. 
In order to confirm this $1/N^2_{~}$ dependence, we show
$[ E_{\rm P}^{N} / N - E_{\rm S}^{N} / B^{N}_{~} ] N^2_{~}$ in Fig.~4, where the value
converges to a constant in the limit $N \rightarrow \infty$. Calculating the
ratio between $| E^N_{\rm P} / N + 2 t / \pi |$ and $| E^N_{\rm S} / B^N_{~} + 2 t / \pi |$,
we find that the former is twice as large as the latter in the limit $N \rightarrow \infty$. 
The result suggests that the spherically deformed $N$-site system is related to a
system of size $2N$ with periodic boundary conditions.

\begin{figure}
\centerline{\includegraphics[width=8.2cm,clip]{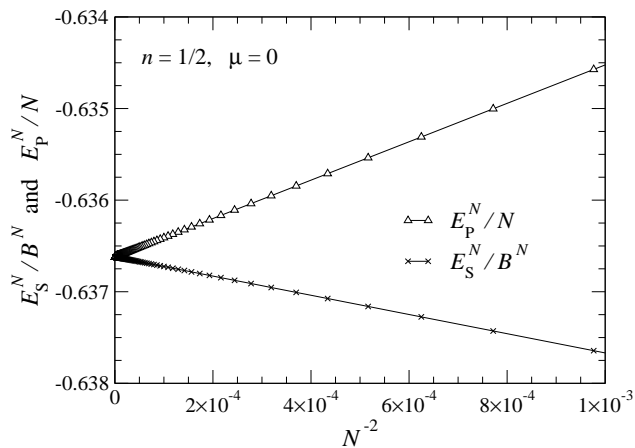}}
\caption{
The finite-size corrections to the energy per site at half filling $n = 1 / 2$.
Crosses show $E^N_{\rm S} / B^N_{~}$ and the open circles $E^N_{\rm P} / N$.
}
\label{f3}
\end{figure}
\begin{figure}
\centerline{\includegraphics[width=8.2cm,clip]{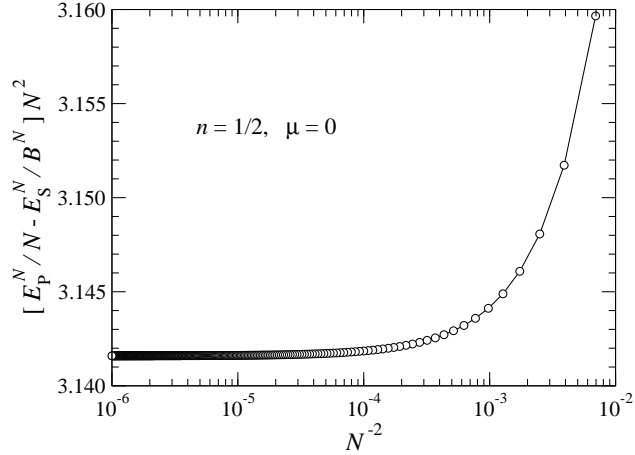}}
\caption{
The convergence of $[ E_{\rm P}^{N} / N - E_{\rm S}^{N} / B^{N}_{~} ] N^2_{~}$ with respect
to $1 / N^2_{~}$.}
\label{f4}
\end{figure}

We have considered the half-filled case. Away of the half filling
we must include the chemical potential term, which is proportional to $\mu$,
into the deformed Hamiltonian. A natural way of introducing $\mu$ is to put it 
into the bond operator $\hat h_{\ell, \ell+1}^{~}$, as stated in Eq.~(1$\cdot$2). 
From this extension we obtain the following Hamiltonian
\begin{equation}
\hat H^{~}_{\rm S} =  \sum_{\ell = 1}^{N-1}
\sin \frac{  \ell \pi }{N}  \, 
\left( 
- t \, \hat c_{\ell}^{\dagger} \hat c_{\ell+1}^{~}  
- t \, \hat c_{\ell+1}^{\dagger} \hat c_{\ell}^{~} 
- \mu \frac{
\hat c_{\ell}^{\dagger} \hat c_{\ell}^{~}  +
\hat c_{\ell+1}^{\dagger} \hat c_{\ell+1}^{~}
}{2}
\right) \, .
\end{equation}
It is also possible to introduce the spherical deformation to the on-site
terms as
\begin{equation}
\hat H'_{\rm S} =  - t \, \sum_{\ell = 1}^{N-1}
\sin \frac{ \ell \pi }{N}  \, 
\left( 
\hat c_{\ell}^{\dagger} \hat c_{\ell+1}^{~}  + 
\, \hat c_{\ell+1}^{\dagger} \hat c_{\ell}^{~} 
\right) 
- \mu 
\sum_{\ell = 1}^{N}
\sin \frac{  ( \ell - \frac{1}{2} ) \pi }{N}  \, 
\hat c_{\ell}^{\dagger} \hat c_{\ell}^{~}  \, ,
\end{equation}
according to the height in Fig.~1 at each site.
Note that both $\hat H^{~}_{\rm S}$ in Eq.~(3$\cdot$4) and $\hat H'_{\rm S}$ 
in Eq.~(3$\cdot$5) give the same 
thermodynamic limit, and that the chemical potential terms do not
commute with the kinetic energy in both cases. This is in
contrast to the {\it undeformed} Hamiltonian
\begin{equation}
{\hat H}_{\rm O}^{~} =  - t \, \sum_{\ell = 1}^{N-1}
\left( 
\hat c_{\ell}^{\dagger} \hat c_{\ell+1}^{~}  + 
\, \hat c_{\ell+1}^{\dagger} \hat c_{\ell}^{~} 
\right) 
- \mu 
\sum_{\ell = 1}^{N}
\hat c_{\ell}^{\dagger} \hat c_{\ell}^{~}  \, ,
\end{equation}
where the chemical potential term is proportional to the total number of particles.
Since we do not know the analytic formulation of one-particle energy of the
deformed Hamiltonians in Eqs.~(3$\cdot$4) and (3$\cdot$5), the relation 
between $\mu$ and the particle filling $n = \sum_\ell^{~} 
\langle {\hat c}_\ell^\dagger {\hat c}_\ell^{~} \rangle / N$ is non-trivial. 
But we are interested in the cases where $N$ is relatively large, therefore it 
is possible to use the relation
$
\mu = - 2t \, \cos( \pi n )
$,
which is satisfied by the undeformed Hamiltonian in Eq.~(3$\cdot$6) in the
limit $N \rightarrow \infty$, as a good approximation
for $\mu$ for the spherically deformed system.

Let us observe the occupation $n_\ell^{~} = 
\langle \hat c_{\ell}^\dagger \hat c_{\ell}^{~} \rangle$ 
with respect to the position $\ell$ at $1/2$, $1/4$ and $1/8$ fillings, respectively, 
where the corresponding $\mu$ is $0$, $- 2 \cos( \pi / 4 )$, and 
$- 2 \cos( \pi / 8 )$. It is obvious that $n_\ell^{~}$ is always $1/2$ at
half filling, equivalently, when $\mu = 0$. Figure 5 shows $n_\ell^{~}$ 
calculated for ${\hat H}_{\rm S}^{~}$ in Eq.~(3$\cdot$4) when $N = 1000$.
The particle distribution is almost uniform, since the ratio of the
hopping strength and the chemical potential is independent on the position $\ell$ on
the lattice. Figure 6 shows $n_\ell^{~}$ near the boundary of the system. The
oscillations in $n_\ell^{~}$ decay rapidly with the distance from the boundary. 
It should be noted that the amplitude of this small oscillation in the particle 
density decreases with increasing the system size $N$.

\begin{figure}
\centerline{\includegraphics[width=8.2cm,clip]{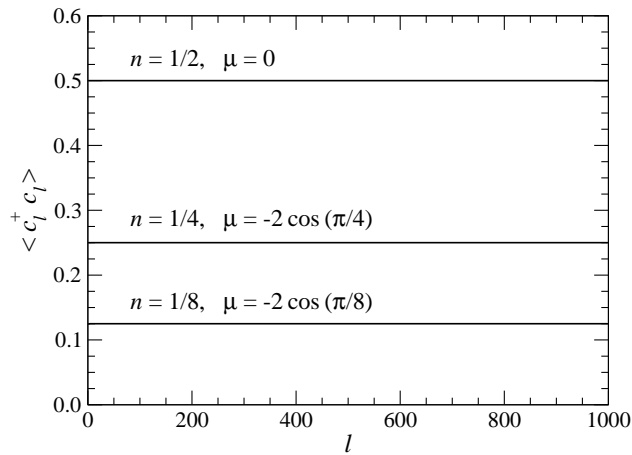}}
\caption{
Occupation number $n_\ell^{~} = 
\langle \hat c_{\ell}^\dagger \hat c_{\ell}^{~} \rangle$ calculated at
$1/2$, $1/4$, and $1/8$ filling, respectively, corresponding to the 
chemical potential  $\mu = 0$, 
$- 2 \cos( \pi / 4 )$, and 
$- 2 \cos( \pi / 8 )$ for ${\hat H}_{\rm S}^{~}$ in Eq.~(3$\cdot$4).}
\label{f5}
\end{figure}
\begin{figure}
\centerline{\includegraphics[width=8.2cm,clip]{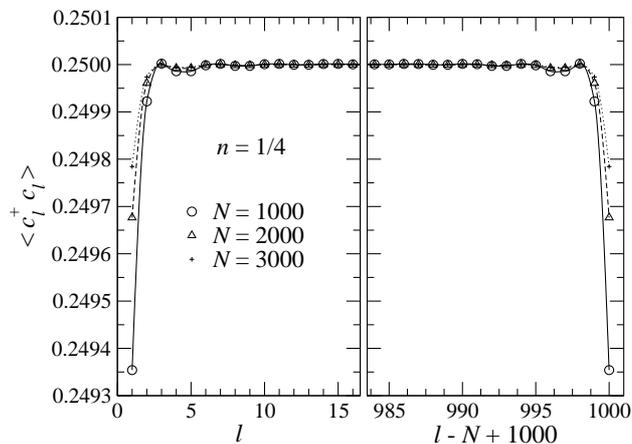}}
\caption{
Position dependence in $n_\ell^{~} = \langle {\hat c}_\ell^\dagger 
{\hat c}_\ell^{~} \rangle$ near the system boundary at quarter filling.}
\label{f6}
\end{figure}

Figure 7 shows the finite-size correction to the energy per bond 
at $1 / 4$ and $1 / 8$ fillings, calculated for both $H_{\rm S}^{~}$ in Eq.~(3$\cdot$4) 
and $H_{\rm S}'$ in Eq.~(3$\cdot$5). As it is observed at half filling shown
in Figs.~3 and 4, the correction is again proportional to $1 / N^2_{~}$. 
We have thus confirmed the $1 / N^2_{~}$ scaling for the correction 
to the ground-state energy per site of the spherically deformed 
lattice-free-Fermion model.

\begin{figure}
\centerline{\includegraphics[width=8.2cm,clip]{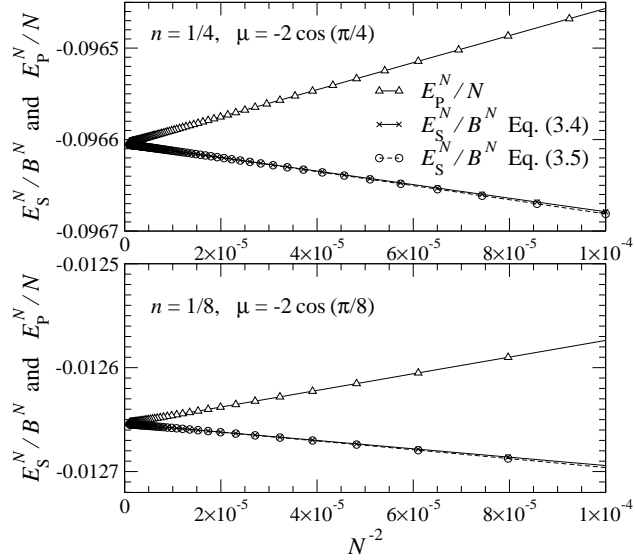}}
\caption{
The finite-size corrections to the energy per site, where the crosses show
$E^N_{\rm S} / B^N_{~}$ calculated from the Hamiltonian $H_{\rm S}^{~}$ Eq.~(3.4), 
the open circles $E^N_{\rm S} / B^N_{~}$ from $H_{\rm S}'$ in Eq.~(3.5). 
For comparison we also show the correction $E^N_{\rm P} / N$ for the
system with periodic boundary conditions by the triangles.}
\label{f7}
\end{figure}

\section{Geometrical interpretation}

There is a 2D classical system behind a 1D quantum system, where the 
relation is called as the quantum-classical correspondence. We show that
spherically deformed Hamiltonian $\hat H_{\rm S}^{~}$ corresponds to
a classical system on a sphere. 
We first consider the quantum-classical correspondence by way of the Trotter
decomposition~\cite{Trotter,Suzuki}. 
For simplicity we consider the half-filled case ($\mu = 0$) for the moment. 
Let us divide $\hat H_{\rm S}^{~}$ in 
Eq.~(3$\cdot$4) into two parts
\begin{equation}
\hat H_{\rm S}^{~} =
\sum_{\ell = {\rm even}}^{~} A_\ell^{~} \, \hat h_{\ell, \ell+1}^{~} +
\sum_{\ell = {\rm odd}}^{~} A_\ell^{~} \, \hat h_{\ell, \ell+1}^{~} =
\hat H_1^{~} + \hat H_2^{~} \, ,
\end{equation}
where we have used the notation 
$h_{\ell,\ell+1}^{~} = - t \left( 
\hat c_{\ell}^{\dagger} \hat c_{\ell+1}^{~} +
\hat c_{\ell+1}^{\dagger} \hat c_{\ell} \right) 
$, 
and where the deformation factor is given by $A_\ell^{~} = \sin ( \ell \pi / N )$.

The imaginary time evolution of amount of $\beta$ is then expressed
by the operator $e^{- \beta \hat H_{\rm S}^{~}}_{~}$. By applying the Trotter
decomposition to $e^{- \beta \hat H_{\rm S}^{~}}_{~}$, we obtain
\begin{equation}
e^{- \beta \hat H_{\rm S}^{~}}_{~} =
\left( e^{- \beta \hat H_{\rm S}^{~} / M}_{~} \right)^M_{~} \sim
\left( e^{- \beta \hat H_1^{~} / M}_{~} \, e^{- \beta \hat H_2^{~} / M}_{~} \right)^M_{~} =
\left( e^{- \Delta \beta \hat H_1^{~}}_{~} \, e^{- \Delta \beta \hat H_2^{~}}_{~} \right)^M_{~} \, ,
\end{equation}
where $M$ is the Trotter number~\cite{Trotter,Suzuki} and $\Delta \beta = 
\beta / M$. Looking at the structure of infinitesimal time evolution by $\hat H_1^{~}$
\begin{equation}
e^{- \Delta \beta \hat  H_1^{~}}_{~} = \exp\left( - \Delta \beta \sum_{\ell =
{\rm even}}^{~} A_\ell^{~} \, \hat h_{\ell, \ell+1}^{~} \right) =
\exp\left( -  \sum_{\ell = {\rm even}}^{~} \left( \Delta \beta A_\ell^{~} \right) \,
\hat h_{\ell, \ell+1}^{~} \right)\, ,
\end{equation}
we find that the quantity
\begin{equation}
\Delta \tau_\ell^{~} = \Delta \beta \, A_\ell^{~}
\end{equation}
plays the role of the rescaled imaginary time. We can treat $e^{-
\Delta \beta \hat H_2^{~}}_{~}$ in the same manner. It is possible to
interpret $\Delta \tau_\ell^{~}$ as a kind of {\it proper time}~\cite{proper} at the
position $\ell$. Such interpretation leads us to an
inhomogeneous time evolution on a (multiply covered) sphere as shown in Fig.~8.
This is the reason why we have used the term {\it spherical
deformation}. Since the surface of the sphere is equivalent everywhere, it is
natural to expect that the ground state of the spherically
deformed Hamiltonian is approximately uniform.

\begin{figure}
\centerline{\includegraphics[width=6.2cm,clip]{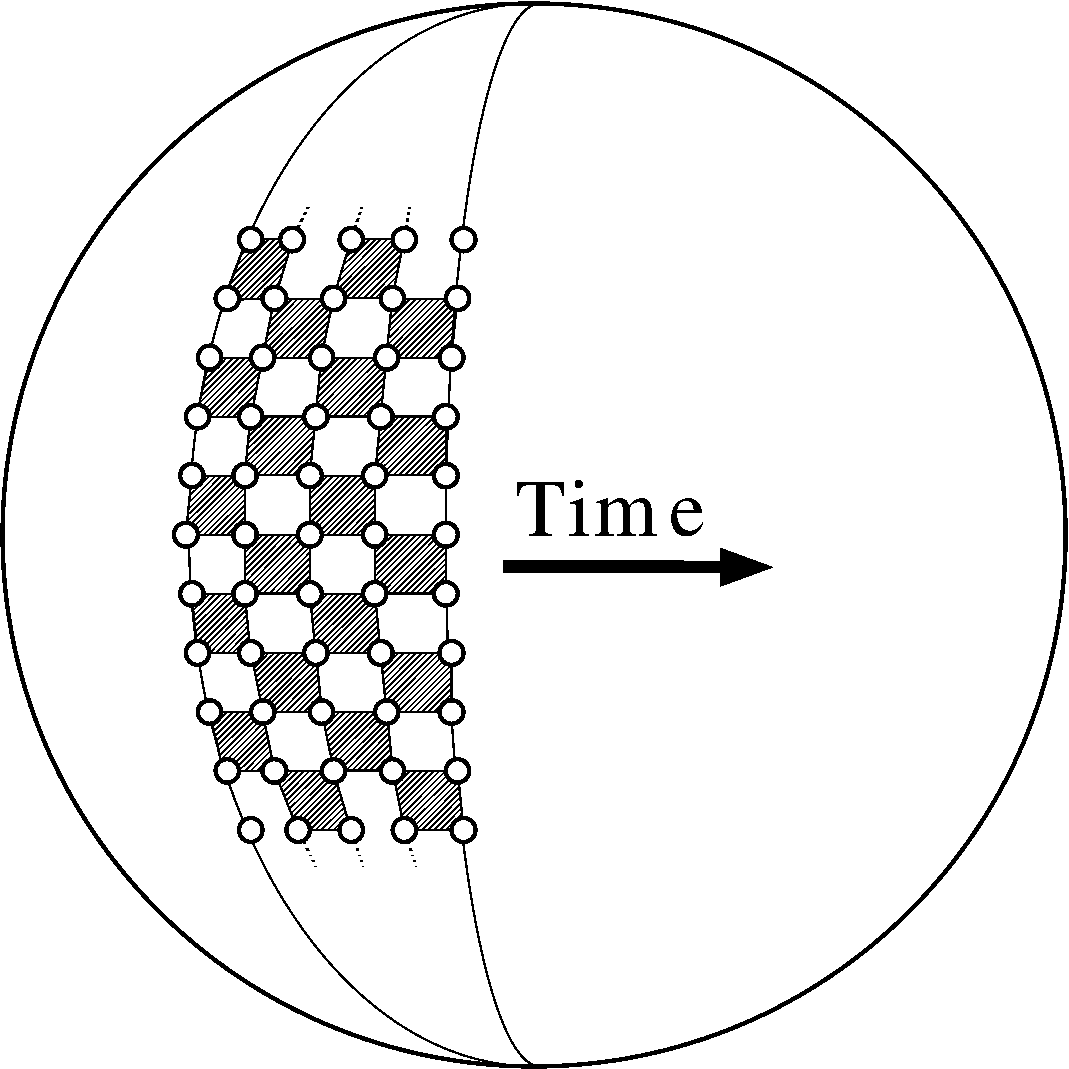}}
\caption{Imaginary time evolution on a sphere.}
\label{f8}
\end{figure}

In the rest of this section we 
show the correspondence with the spherical geometry by taking 
the continuous limit to the lattice Hamiltonian $\hat H_{\rm S}^{~}$ or
${\hat H}'_{\rm S}$. 
Consider a 1-particle state
\begin{equation}
| \psi( {\tt t} )\rangle = \sum_{\ell = 1}^{N} 
\psi_{\ell}^{~}( {\tt t} ) \, \hat c_\ell^\dagger | 0 \rangle 
\end{equation}
at time ${\tt t}$. (Since we have been using the letter $t$ for the hopping
parameter, we use ${\tt t}$ for the time.) 
The real-time evolution of the wave function $\psi_\ell^{~}( {\tt t} )$ is
described by the Schr\"odinger equation 
\begin{equation}
i \hbar \frac{\partial}{\partial {\tt t}} \psi_\ell^{~} 
=
- t \, \sin \frac{ \ell \pi }{N}  \, \psi_{\ell+1}^{~}
- t \, \sin \frac{ ( \ell - 1 ) \pi }{N} \, \psi_{\ell-1}^{~}
- \mu 
\sin  \frac{ ( \ell - \frac{1}{2} ) \pi }{N} \, \psi_\ell^{~} \, 
\end{equation}
under the Hamiltonian $\hat H_S'$ in Eq.~(3$\cdot$5). Note that the difference between 
$\hat H_S^{~}$ in Eq.~(3$\cdot$4) and $\hat H_S'$ in Eq.~(3$\cdot$5) is not 
relevant in the large $N$ limit. There are two different continuous limits for this spatially 
discrete Schr\"odinger equation. We first consider the massive case where $\mu$ is
nearly equal to $- 2t$. Introducing the notation 
$f_\ell^{~} = \sin\left[ ( \ell - \frac{1}{2} ) \pi / N  \right]$, we
can rewrite Eq.~(4$\cdot$6) by use of differentials
\begin{eqnarray}
i \hbar \frac{\partial}{\partial {\tt t}} \psi_\ell^{~} 
&=&
- t \, f_{\ell + \frac{1}{2}}^{~} \, \psi_{\ell+1}^{~} 
- t \, f_{\ell - \frac{1}{2}}^{~} \, \psi_{\ell - 1}^{~} 
- \mu \, f_\ell^{~} \, \psi_\ell^{~} \\
&=&
- t \left[ 
f_{\ell + \frac{1}{2}}^{~} \left( \psi_{\ell+1}^{~} - \psi_\ell^{~} \right) -
f_{\ell - \frac{1}{2}}^{~} \left( \psi_\ell^{~} - \psi_{\ell - 1}^{~} \right)
\right]
- \left(
\mu \, f_\ell^{~} + t \, f_{\ell+\frac{1}{2}}^{~} + t \, f_{\ell - \frac{1}{2}}^{~}
\right) \psi_\ell^{~} \, , \nonumber
\end{eqnarray}
where we have substituted the trivial relations 
 $\psi_{\ell + 1}^{~} = ( \psi_{\ell + 1}^{~} - \psi_{\ell}^{~} ) + \psi_{\ell}^{~}$ and 
 $\psi_{\ell - 1}^{~} = - ( \psi_{\ell}^{~} - \psi_{\ell - 1}^{~} ) + \psi_{\ell}^{~}$.
Using the relations
\begin{eqnarray}
f_{\ell + \frac{1}{2}}^{~} &=& \frac{1}{2}
\left( f_{\ell + \frac{1}{2}}^{~} + f_{\ell - \frac{1}{2}}^{~} \right) +
\left( f_{\ell + \frac{1}{2}}^{~} - f_{\ell - \frac{1}{2}}^{~} \right) \nonumber\\
f_{\ell - \frac{1}{2}}^{~} &=& \frac{1}{2}
\left( f_{\ell + \frac{1}{2}}^{~} + f_{\ell - \frac{1}{2}}^{~} \right) -
\left( f_{\ell + \frac{1}{2}}^{~} - f_{\ell - \frac{1}{2}}^{~} \right)
\end{eqnarray}
we can further rewrite Eq.~(4$\cdot$7) as
\begin{eqnarray}
i \hbar \frac{\partial}{\partial {\tt t}} \psi_\ell^{~} = 
&-&
\frac{t}{2}\left( f_{\ell + \frac{1}{2}}^{~} + f_{\ell - \frac{1}{2}}^{~} \right)
\left[ 
\left( \psi_{\ell + 1}^{~} - \psi_\ell^{~} \right) - 
\left( \psi_\ell^{~} - \psi_{\ell - 1}^{~} \right) \right] \nonumber\\
&-&
\frac{t}{2}\left( f_{\ell + \frac{1}{2}}^{~} - f_{\ell - \frac{1}{2}}^{~} \right)
\left[ 
\left( \psi_{\ell + 1}^{~} - \psi_\ell^{~} \right) + 
\left( \psi_\ell^{~} - \psi_{\ell - 1}^{~} \right) \right] \nonumber\\
&-& 
\frac{t}{2}\left( f_{\ell + \frac{1}{2}}^{~} + f_{\ell - \frac{1}{2}}^{~} \right)
2 \, \psi_\ell^{~} - \mu  \, f_\ell^{~} \psi_\ell^{~} \, .
\end{eqnarray}

Now we introduce the lattice constant $a = \pi R / N$, where 
$R$ is the radius of the sphere. We also introduce the spacial 
co-ordinate $x = a ( \ell - \frac{1}{2} )$, which satisfies $0 < x < \pi R$. 
Using these notations we rewrite 
$\psi_\ell^{~}( {\tt t} )$ as $\psi( x, {\tt t} )$, and
$f_\ell^{~}$ as $f( x ) = \sin (x / R) = \sin \theta$, where 
$\theta = x / R$ is {\it the angle measured from the north pole}. The continuous limit can be
taken by increasing the number of sites $N$ keeping $R$ constant, 
where the lattice constant $a$ decreases with $N$.
Simultaneously we increase the hopping parameter $t$ so that the relation 
$a^2_{~} t = \hbar^2_{~} / 2m$ always holds, where $m$ is the particle
mass, and $\hbar$ the Dirac constant. To prevent the divergence in the
potential term, we adjust $\mu$ so that $\mu + 2t = - V$ is satisfied, where
$V$ is a finite constant. Using these parametrizations, we obtain the
Schr\"odinger equation in continuous space
\begin{equation}
i \hbar \frac{\partial}{\partial {\tt t}} \psi( x, {\tt t} ) = - \frac{\hbar^2_{~}}{2m}
\frac{\partial}{\partial x} \left[ f( x ) \frac{\partial}{\partial x} \psi( x, {\tt t} ) \right] + 
V \, f( x ) \psi( x, {\tt t} ) \, .
\end{equation}
This equation is derived from the Lagrangian
\begin{equation}
{\cal L} = - i \hbar \, \psi^*_{~}( x, {\tt t} ) \frac{\partial}{\partial {\tt t}} \psi( x, {\tt t} ) +
f( x ) \left[ \frac{\hbar^2_{~}}{2m} 
\frac{\partial \psi^*_{~}( x, {\tt t} )}{\partial x}
\frac{\partial \psi( x, {\tt t} )}{\partial x} + V \psi^*_{~}( x, {\tt t} ) \psi( x, {\tt t} )
\right] \, ,
\end{equation}
where introduction of proper time $\tau( x, {\tt t} )$ that satisfies
${\rm d} {\tt t} = \frac{1}{f( x )} {\rm d} \tau( x, {\tt t} )$ draws the following Lagrangian
\begin{equation}
{\cal L} = f( x ) \left[ - i \hbar \, \psi^*_{~}( x, \tau ) \frac{\partial}{\partial \tau} \psi( x, \tau ) + 
\frac{\hbar^2_{~}}{2m} 
\frac{\partial \psi^*_{~}( x, \tau )}{\partial x}
\frac{\partial \psi( x, \tau )}{\partial x} + V \psi^*_{~}( x, \tau ) \psi( x, \tau )
\right] 
\end{equation}
in the $x$-$\tau$ space. The action ${\cal S}$ is then written as
\begin{equation}
{\cal S} = \int  \left[
- i \hbar \, \psi^*_{~} \frac{\partial}{\partial \tau} \psi + 
\frac{\hbar^2_{~}}{2m} 
\frac{\partial \psi^*_{~}}{\partial x}
\frac{\partial \psi}{\partial x} + V \psi^*_{~} \psi
\right] \sin \frac{x}{R} \,\,  {\rm d} \tau  {\rm d} x \, .
\end{equation}
As it is seen, $f( x ) \, {\rm d} \tau {\rm d} x = \sin( x / R ) \,  {\rm d} \tau {\rm d} x$ 
plays the role of the integral measure 
on the sphere of radius $R$. Note that the continuous limit for the field
operator $\hat c_\ell^{~} \rightarrow \hat \psi( x )$ can be taken in the same
manner as in Eqs.~(4$\cdot$6)-(4$\cdot$13) using the correspondence 
in Eq.~(4$\cdot$5).

Since $f( x ) = 0$ at the both ends, where $x = 0$ and $x = \pi R$, the 
continuous one-dimensional quantum system in Eqs.~(4$\cdot$10)-(4$\cdot$13) 
does not effectively contain the system boundaries.
We can observe the fact by way of the conformal mapping
\begin{equation}
y = - R \, \log \, \cot \left( \frac{x}{2R} \right) 
\end{equation}
from the sphere embedded in three dimensions onto the infinite plane. We have the relations
%
%
%
%
\begin{equation}
\sin\frac{x}{R} = 2 \sin\frac{x}{2R} \cos\frac{x}{2R} = \left( \cosh\frac{y}{R} \right)^{-1}_{~}
\end{equation}
and $\sin ( x / R ) \, dy = \sin \theta \, dy = dx$. 
The action $\cal S$ on this infinite ${\tt t}$-$y$ plane is then written as 
\begin{equation}
{\cal S} = \int  \left[
- i \hbar \, \psi^*_{~} \left( \frac{1}{\sin \theta} \frac{\partial \psi}{\partial {\tt t}}  \right) + 
\frac{\hbar^2_{~}}{2m} 
\left( \frac{1}{\sin \theta} \frac{\partial \psi^*_{~}}{\partial y} \right)
\left( \frac{1}{\sin \theta} \frac{\partial \psi}{\partial y} \right) + 
V \psi^*_{~} \psi
\right] \sin^2_{~} \theta \,\,  {\rm d} {\tt t}  {\rm d} y \, ,
\end{equation}
where $\sin \theta = \left( \cosh \frac{y}{R} \right)^{-1}_{~}$ is satisfied.
%
%
The corresponding one-particle Hamiltonian is obtained as follows
\begin{equation}
H =
- \frac{\hbar^2_{~}}{2m} \frac{\partial}{\partial x}
\left( \sin\frac{x}{R} \frac{\partial}{\partial x} \right) + V \, \sin\frac{x}{R}
= 
- \frac{\hbar^2_{~}}{2m} \cosh\frac{y}{R} \frac{\partial^2_{~}}{\partial y^2_{~}}
+ V \left( \cosh\frac{y}{R} \right)^{-1}_{~} \, .
\end{equation}

We can also formulate a massless limit, which appears in 
the case $- 2 t < \mu < 2t$ where there is a Fermi surface,
in the same manner as in Eqs.~(4$\cdot$6)-(4$\cdot$9).
In this case we substitute 
$\psi_{\ell}^{~} = e^{\pm i k \ell}_{~} \phi_{\ell}^{~}$ to Eq.~(4$\cdot$6), 
where $k$ and $- k$ are, respectively, the Fermi wave number for 
the right and the left going modes. 
One finds that the quantity $\nu = a t$ is the leading order
in the small lattice constant limit $a \rightarrow 0$, equivalently in the
large $N$ limit. Adjusting $\mu$ so that
$2 t \cos k + \mu = -V$ is satisfied, we obtain the equation of motion 
\begin{equation}
i \hbar \frac{\partial}{\partial \tt t} \phi( x, {\tt t} ) = \mp 2 i \nu \sin k 
\frac{\partial}{\partial x} \bigl[ f( x ) \phi( x, {\tt t} ) \bigr] + V f( x ) \phi( x, {\tt t} ) \, ,
\end{equation}
for the continuous field $\phi( x, {\tt t} )$.
The corresponding Lagrangian in $x$-$\tau$ plane is 
\begin{equation}
{\cal L} = f( x ) \left[ 
- i \hbar \, \phi^*_{~}( x, \tau ) \frac{\partial}{\partial \tau} \phi( x, \tau ) \pm 
2 i \nu \sin k \, \phi( x, \tau ) 
\frac{\partial \phi^*_{~}( x, \tau )}{\partial x} + V \psi^*_{~}( x, \tau ) \psi( x, \tau )
\right] \, ,
\end{equation}
where we have used the fact that $f( 0 ) = f( 2R) = 0$. Similar to
Equations (4$\cdot$13)-(4$\cdot$16), we can consider the conformal 
mapping for this massless case. 
The $1/N^2_{~}$ dependence of the corrections to the ground-state energy 
per site might be explained by the boundary conformal field theory, 
where we leave the conjectures for the future study.

\section{Conclusions and discussions}

We have investigated the ground state of the spherically deformed
1D free Fermion system, for both at the half filling and away of the half filling. 
The finite-size correction to the energy per site is of the order of $1 /
N^2_{~}$ for both cases. The reason for such fast convergence is qualitatively
explained by the quantum-classical correspondence, where the
spherically deformed Hamiltonians essentially correspond to classical
fields on a sphere. In such a sense the spherically deformed system does
not contain the system boundary.

Interest in the spherical deformation rests in dynamical properties.
We conjecture that a moving one-particle wave packet on
the spherically deformed lattice oscillates nearly harmonically as a consequence of the
circulation on the sphere. The oscillation may be also explained
by a continuous refraction caused by a slower dynamics near the
both ends of the system.

As a generalizations of the spherically deformed Hamiltonian ${\hat H}_{\rm S}^{~}$
in Eq.~(1$\cdot$7), one can consider a decoupled Hamiltonian
\begin{equation}
\hat H_{\rm sin}^{~} = \sum_{\ell = 1}^{2N-1} 
\sin \frac{\ell \pi}{N} \, 
\left( \hat h_{\ell, \ell+1}^{~} + \frac{ \hat g_\ell^{~} + \hat g_{\ell+1}^{~}}{2} \right)  \, ,
\end{equation}
where $\ell = 2N + 1$ is equivalent to $\ell = 1$, for a system of size $2N$.
The differential with respect to $\ell$ draws
\begin{equation}
\hat H_{\rm cos}^{~} = \sum_{\ell = 1}^{2N-1} 
\cos \frac{\ell \pi}{N} \, 
\left( \hat h_{\ell, \ell+1}^{~} + \frac{ \hat g_\ell^{~} + \hat g_{\ell+1}^{~}}{2} \right)  \, ,
\end{equation}
which is again the decoupled Hamiltonian when $N$ is an even number.~\cite{hyp}
Both $\hat H_{\rm sin}^{~}$ and $\hat H_{\rm cos}^{~}$ seems to be 
generators of rotation on a kind of discrete sphere. Their commutation relation
would be discussed elsewhere.

If one is interested in the estimation of the excitation gap, the
spherical deformation is not appropriate. This is because weak bonds
near the system boundary induce spurious low-energy excitations.
For this purpose, the hyperbolic deformation is more
appropriate~\cite{hyp,hyp2}. The quantum-classical correspondence
discussed in this article can be also considered for the hyperbolic deformation, 
which would deduce continuous field model on the Poincare disc.

\section*{Acknowledgements}

The authors thank to U.~Schollw\"ock for stimulating
discussions and encouragement. T.~N. is grateful to K.~Okunishi for
valuable discussions about deformations. T.~N. thank to G.~Sierra for
valuable comments. This work is partially supported by
Slovak Agency for Science and Research grant APVV-51-003505,
APVV-VVCE-0058-07, QUTE, and VEGA grant No. 1/0633/09 (A.G. and R.K.) as well as
partially by a Grant-in-Aid for Scientific Research from Japanese
Ministry of Education, Culture, Sports, Science and Technology
(T.N. and A.G.). A.G. acknowledges support of the Alexander von
Humboldt foundation.

\newpage
\begin{figure}[!htp]
\centerline{\includegraphics[height=21.0cm,clip]{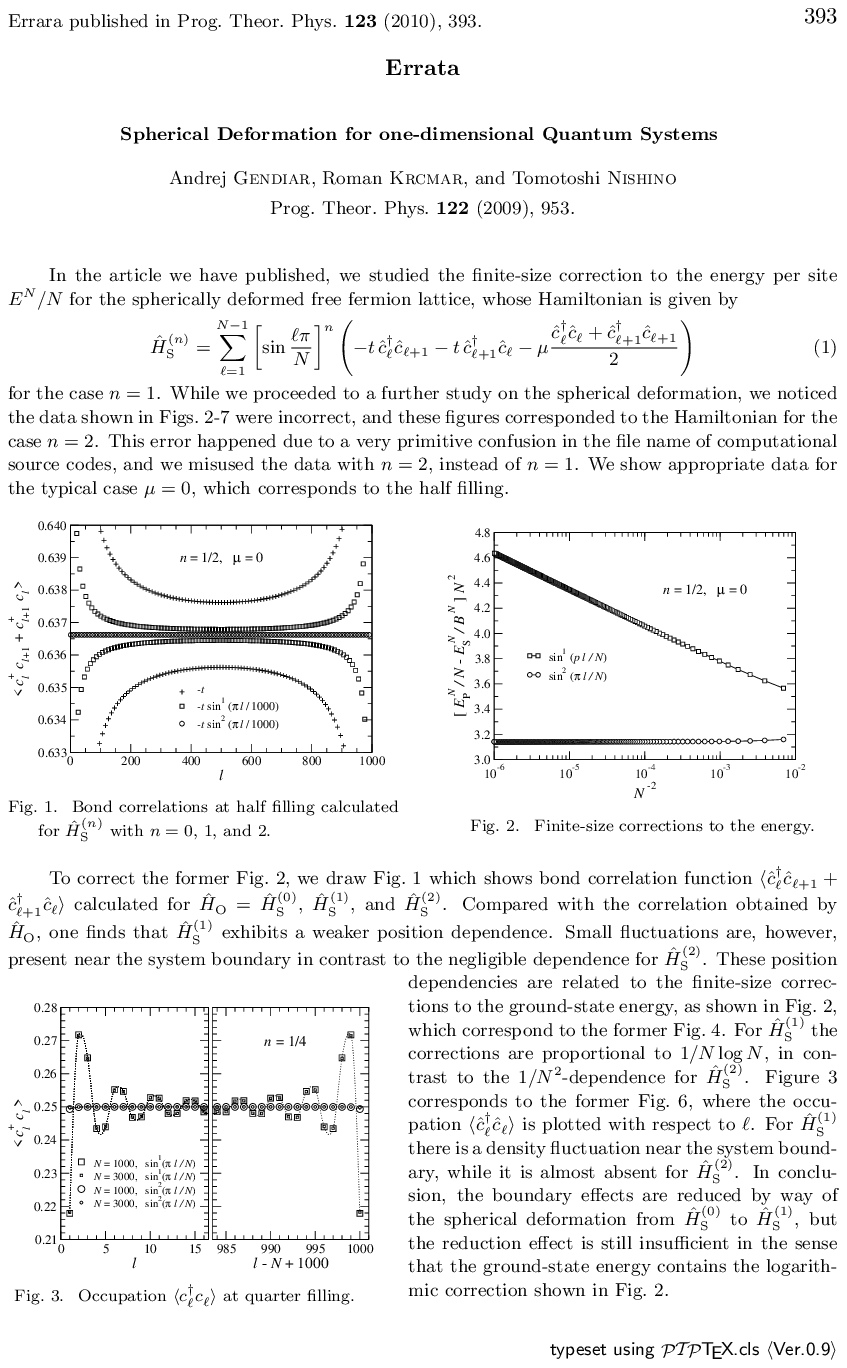}}
\end{figure}
\thispagestyle{empty}


\begin{thebibliography}{99}
\bibitem{proper} Note that we are considering an infinitesimal time evolution, and 
we do not consider the global geodesics that starts from the position $\ell$ on the 
discretized sphere.
\end{thebibliography}
\end{document}